\documentstyle[prl,aps]{revtex}
\large  

\input epsf
\begin{document}
\title{Sign-time distribution for a random walker with a drifting boundary} 
\author{T. J. Newman} 
\address{Departments of Physics and Biology, University of Virginia,
Charlottesville, VA 22904}
\maketitle
\begin{abstract}
We present a derivation of the exact sign-time distribution for
a random walker in the presence of a boundary moving with constant velocity.
\end{abstract}
\vspace{0.5cm}

There has been tremendous interest within the statistical physics community
over the past five years in the persistence properties of various simple
systems\cite{rev}. Aside from the persistence probability itself, 
another quantity of great interest is the ``sign-time'' distribution 
(STD)\cite{std1,std2} (also known as the ``occupation-time'' or 
``residence-time'' distribution). It seems
that STD's are very difficult to calculate, even for the simplest
systems. Famous examples for which the STD is known exactly are the pure
random walk, yielding the arcsine law of Levy\cite{levy,feller}, and the 
renewal process, originally studied by Lamperti\cite{lamp}, and recently
revisited\cite{bald,gl}. There have been concerted efforts recently to find
the STD for a generalized random walk model\cite{dhar,desm}, which have
met with some success, although an exact solution is still not known. 
Here we consider the STD for a random walker
in the presence of a boundary moving with constant velocity (which is 
equivalent to a random walker with a constant drift velocity in the presence 
of a stationary boundary). The first passage
probability\cite{bl} and the persistence probability\cite{bcl} for this
problem are both known. However, to the author's knowledge, the STD has 
not been previously calculated. The purpose of this paper is to present 
a relatively straightforward derivation and analysis of this latter quantity.

We consider a random walker whose position is denoted by $x(t)$, which 
satisfies the equation of motion
\begin{equation}
\label{model}
{\dot x}=\xi(t) \ ,
\end{equation}
with $x(0)=0$, and $\langle \xi(t) \rangle = 0$ and 
$\langle \xi(t)\xi(t') \rangle
=D\delta (t-t')$. In addition we consider a moving boundary whose position is
denoted by $b(t)$. In the main part of this paper we shall consider
linear motion of the boundary, i.e. $b(t)=vt$.

The sign-time $\tau _{\xi}(t)$ is defined as the proportion of elapsed
time $t$ the walker has spent on the positive side of the boundary:
\begin{equation}
\label{sign-time}
\tau _{\xi}(t) = \int \limits _{0}^{t} dt' \ \Theta (x(t')-b(t')) \ ,
\end{equation}
where $\Theta (x)$ is the step function. We are interested in
calculating the STD $P(\tau, t)$ defined by
\begin{equation}
\label{std}
P(\tau,t) = \left \langle \delta (\tau - \tau _{\xi}(t)) \right \rangle \ .
\end{equation}
By making an integral representation of $P(\tau,t)$, we can reduce the problem
to a calculation of the moments of the STD:
\begin{eqnarray}
\label{std-mom1}
P(\tau,t) & = & \int \limits _{-\infty}^{\infty} {d\omega \over 2\pi} \
e^{-i\omega \tau} \ \left \langle \exp \left [i\omega \tau _{\xi}(t) \right ]
\right \rangle \ , \\
\label{std-mom2}
& = & \int \limits _{-\infty}^{\infty} {d\omega \over 2\pi} \ e^{-i\omega \tau}
\ \sum \limits _{n=0}^{\infty} {(i\omega)^{n} \over n! } 
\left \langle \tau _{\xi}(t)^{n} \right \rangle \ .
\end{eqnarray}

To this end we introduce a convenient double-integral representation of
the step function
\begin{equation}
\label{stepfn}
\Theta (x) = \int \limits _{0}^{\infty} du \ \delta (x-u) 
= \int \limits _{0}^{\infty} du \ \int \limits _{-\infty}^{\infty}
{dk \over 2\pi} \ e^{ik(x-u)} \ .
\end{equation}
Utilizing this in the definition of the sign-time, we can write the 
$n^{\rm th}$ moment as the following time-ordered expression:
\begin{eqnarray}
\label{nthmom1a}
\left \langle \tau _{\xi}(t)^{n} \right \rangle & = &
n! \int{\cal D}t(n) \left \langle \Theta (x(t_{1})-b(t_{1}))
\cdots \Theta (x(t_{n})-b(t_{n})) \ \right \rangle \ , \\
\label{nthmom1b}
& = & n! \int {\cal D}t(n) \int {\cal D}u(n) \int {\cal D}k(n)
\ e^{-ik_{1}(u_{1}+b(t_{1}))}\cdots 
e^{-ik_{n}(u_{n}+b(t_{n}))} \ \left \langle e^{ik_{1}x(t_{1})}
\cdots e^{ik_{n}x(t_{n})} \right \rangle \ ,
\end{eqnarray}
where the multiple integral symbols are defined via
\begin{eqnarray}
\label{nthmom2}
\int{\cal D}t(n) & \equiv & \int \limits _{0}^{t} dt_{1} 
\int \limits _{0}^{t_{1}} dt_{2} \cdots 
\int \limits _{0}^{t_{n-1}} dt_{n} \ , \\
\int{\cal D}u(n) & \equiv & \int \limits _{0}^{\infty} du_{1}
\cdots \int \limits _{0}^{\infty} du_{n} \ , \\
\int{\cal D}k(n) & \equiv & \int \limits _{-\infty}^{\infty} {dk_{1}\over 2\pi}
\cdots \int \limits _{-\infty}^{\infty} {dk_{n}\over 2\pi} \ .
\end{eqnarray}
Writing the solution of Eq.(\ref{model}) as $x(t)=\int \limits _{0}^{t}
dt'\xi(t')$, the average on the right-hand-side of Eq.(\ref{nthmom1b}) 
is easily performed to give
\begin{equation}
\label{expav}
\left \langle e^{ik_{1}x(t_{1})} \cdots e^{ik_{n}x(t_{n})} \right \rangle
= \exp \left \lbrace -{D \over 2} \left [ k_{1}^{2}(t_{1}-t_{2})
+ (k_{1}+k_{2})^{2}(t_{2}-t_{3}) + \cdots + (k_{1}+\cdots+k_{n})^{2}t_{n}
\right ] \right \rbrace \ .
\end{equation}
Inserting this expression into Eq.(\ref{nthmom1b}) and
changing variables via $(p_{1}=k_{1}, \dots, p_{n}=k_{1}+\cdots+k_{n})$, we 
find the integrals over $\lbrace p_{i}\rbrace $ decompose and may be
performed, giving us
\begin{equation}
\label{nthmom3}
\left \langle \tau _{\xi}(t)^{n} \right \rangle =
n! \int {\cal D}t(n) \int {\cal D}u(n) 
\ G(u_{1}+b(t_{1})-u_{2}-b(t_{2}),t_{1}-t_{2})
\cdots G(u_{n}+b(t_{n}),t_{n}) \ ,
\end{equation}
where $G(u,t)$ is the random walk propagator
\begin{equation}
\label{prop}
G(u,t) = (2\pi Dt)^{-1/2}\exp [-u^{2}/2Dt] \ .
\end{equation}

We shall now focus on the case of a boundary drifting with constant
velocity $v$, i.e. we write $b(t)=vt$. We note the appearance of a fundamental 
rate constant $\gamma \equiv v^{2}/2D$. It is convenient to 
introduce the function $H(u,t)=G(u+vt,t)$. By considering the Laplace
transform of the $n^{\rm th}$ moment, and invoking the Laplace transform
convolution theorem iteratively, we find
\begin{equation}
\label{nthmom4}
{\cal L}_{s|t} \left [ \left \langle \tau _{\xi}(t)^{n} \right \rangle 
\right ] = {n! \over s} \int {\cal D}u(n) \ {\hat H}(u_{1}-u_{2},s)
\cdots {\hat H}(u_{n-1}-u_{n},s){\hat H}(u_{n},s) \ ,
\end{equation}
where the Laplace transform operator is denoted by 
\begin{equation}
\label{ltdef}
{\hat f}(s) = {\cal L}_{s|t} [f(t)] = \int \limits _{0}^{\infty} 
dt \ e^{-st}f(t) \ . 
\end{equation}

Now, $H(u,t)=\exp(-uv/D-\gamma t)G(u,t)$ and consequently
${\hat H}(u,s) = e^{-uv/D}{\hat G}(u,s+\gamma)$. The Laplace transform
of $G(u,t)$ is given by
\begin{equation}
\label{proplt}
{\hat G}(u,s) = (2sD)^{-1/2}\exp [-(2s/D)^{1/2}|u|] \ .
\end{equation}
Substituting the explicit form for ${\hat H}(u,s)$ into Eq.(\ref{nthmom4})
and rescaling the integration variables we find
\begin{equation}
\label{nthmom5}
{\cal L}_{s|t} \left [ \left \langle \tau _{\xi}(t)^{n} \right \rangle 
\right ] = {n! \over s[2(s+\gamma )]^{n}} 
\int {\cal D}u(n) \ e^{-\alpha (s)u_{1}} \ 
\exp \left (-|u_{1}-u_{2}|-\cdots-|u_{n-1}-u_{n}|-|u_{n}| \right ) \ ,
\end{equation}
where we have defined $\alpha (s) = (\gamma/(s+\gamma))^{1/2}$.
Eq.(\ref{nthmom5}) can be conveniently rewritten, for $n \ge 1$, as
\begin{equation}
\label{nthmom6}
{\cal L}_{s|t} \left [ \left \langle \tau _{\xi}(t)^{n} \right \rangle 
\right ] = {n! \over s[2(s+\gamma )]^{n}} 
\int \limits _{0}^{\infty} du_{1} \ e^{-\alpha (s)u_{1}} J_{n-1}(u_{1}) \ ,
\end{equation}
where the function $J_{n}(u)$ satisfies the integro-difference equation
\begin{equation}
\label{intdiff}
J_{n}(u) = \int \limits _{0}^{\infty} du' \ e^{-|u-u'|} J_{n-1}(u') \ ,
\end{equation}
with $J_{0}=e^{-u}$. This equation for $J_{n}(u)$ can be solved using  
generating function methods, and the details are relegated to an Appendix.
The end result is
\begin{equation}
\label{jnsol}
J_{n}(u) = \oint \limits _{C} {dz \over 2\pi i} \ \left ( {1-(1-2z)^{1/2}\over 
z^{n+2}} \right ) \ \exp [-(1-2z)^{1/2}u] \ ,
\end{equation}
where the contour $C$ encircles the origin. 

We now substitute this solution for $J_{n}(u)$ into Eq.(\ref{nthmom6}) and
perform the integral over $u_{1}$. Using the fact that 
$\oint _{C} dz/z^{n+1}=0$ for $n\ge 1$ allows us to rewrite the
integrand so as to obtain
\begin{equation}
\label{nthmom7}
{\cal L}_{s|t} \left [ \left \langle \tau _{\xi}(t)^{n} \right \rangle 
\right ] = {n! (1+\alpha (s)) \over s[2(s+\gamma )]^{n}} 
\oint \limits _{C} {dz \over 2\pi i} \ {1\over 
z^{n+1}[\alpha (s)+(1-2z)^{1/2}]}  \ .
\end{equation}
The integral around the contour $C$ may be re-expressed as an integral across
the cut emanating from the branch point at $z=1/2$ with the result
\begin{equation}
\label{nthmom8}
{\cal L}_{s|t} \left [ \left \langle \tau _{\xi}(t)^{n} \right \rangle 
\right ] = {n! (1+\alpha (s)) \over \pi s(s+\gamma )^{n}} 
\int \limits _{0}^{\infty} dq \ {q^{1/2} \ 
\over (1+q)^{n+1}(\alpha (s)^{2}+q)} \ .
\end{equation}
It can be explicitly checked that this expression is also correct for
$n=0$, namely, that it gives the result $1/s$.

It is possible to invert the Laplace transform at this point, but it is
more convenient to postpone this operation and instead introduce the
integral representation for $n!$ via
\begin{equation}
\label{nfac}
n!= \int \limits _{0}^{\infty} dy \ y^{n}e^{-y} \ .
\end{equation}
Then Eq.(\ref{nthmom8}) takes the form
\begin{equation}
\label{nthmom9}
{\cal L}_{s|t} \left [ \left \langle \tau _{\xi}(t)^{n} \right \rangle 
\right ] = {(1+\alpha (s)) \over \pi s} 
\int \limits _{0}^{\infty} dq \ {q^{1/2} \ 
\over (1+q)(\alpha (s)^{2}+q)} \int \limits _{0}^{\infty} dy \ e^{-y}
\ \left [ {y \over (s+\gamma)(1+q)} \right ]^{n} \ .
\end{equation}

Referring to Eq.(\ref{std-mom2}) we have
\begin{equation}
\label{stdlt1}
{\cal L}_{s|t}\left [ P(\tau,t) \right ] = 
\int \limits _{-\infty}^{\infty} {d\omega \over 2\pi} \ e^{-i\omega \tau}
\ \sum \limits _{n=0}^{\infty} {(i\omega)^{n} \over n! } 
{\cal L}_{s|t} \left [ \left \langle \tau _{\xi}(t)^{n} \right \rangle 
\right ] \ .
\end{equation}
Substituting Eq.(\ref{nthmom9}) into Eq.(\ref{stdlt1}), we can explicitly
perform the sum over $n$ which allows us to complete the integral over
$\omega $ yielding a Dirac $\delta$-function with argument 
$(\tau -y/(s+\gamma)(1+q))$, which in turn allows us to trivially perform the
integral over $y$. This leaves us with (after rescaling the $q$ integration 
variable),
\begin{equation}
\label{stdlt2}
{\cal L}_{s|t}\left [ P(\tau,t) \right ] = {\alpha (s)(1+\alpha (s))(s+\gamma) 
\over \pi s} 
e^{-(s+\gamma)\tau} \ \int \limits _{0}^{\infty} dq \ {q^{1/2}
e^{-\gamma \tau q} \over (1+q)} \ .
\end{equation}
It is worth noting the clean disappearance of the Laplace transform 
variable $s$ from inside the $q$-integral, which allows us to complete
the calculation easily from this point. Formally inverting the Laplace 
transform gives us
\begin{equation}
\label{stdltinv}
P(\tau,t) =  N(\gamma \tau ) {\cal L}^{-1}_{t|s}\left [ 
{\alpha (s)(1+\alpha (s))(s+\gamma) \over s} e^{-s\tau} \right ] \ ,
\end{equation}
where we have defined
\begin{equation}
\label{finalfn}
N(\beta ) \equiv {e^{-\beta}\over \pi}  
\int \limits _{0}^{\infty} dq \ 
{q^{1/2} e^{-\beta q} \over (1+q)} = {e^{-\beta }\over (\pi \beta )^{1/2}}
-{\rm erfc}(\beta ^{1/2}) \ ,
\end{equation}
with ${\rm erfc}(z)$ the complementary error function\cite{as}.
On performing the inverse Laplace transform in Eq.(\ref{stdltinv}) we obtain
\begin{equation}
\label{stdfinal1}
P(\tau,t) =  \gamma N(\gamma \tau )[2+N(\gamma(t-\tau))] \ .
\end{equation}
This may be cast into a more symmetrical form by defining the function $F^{+}$
and its companion $F^{-}$ via
\begin{equation}
\label{symfns}
F^{\pm}(\beta ) = e^{-\beta }\mp(\pi \beta)^{1/2}{\rm erfc}
(\pm\beta ^{1/2}) \ .
\end{equation}
We then have our final result in the form
\begin{equation}
\label{stdfinal2}
P(\tau,t) =  {F^{+}(\gamma \tau )F^{-}(\gamma(t-\tau)) \over 
\pi (\tau (t-\tau))^{1/2}} \ .
\end{equation}
It is interesting to note that the distribution is a product of a function
of $\tau$ and a function of $(t-\tau)$.
It is often convenient to express the STD in dimensionless variables. 
Defining $\phi = \tau/t \in (0,1)$ and $\eta = \gamma t$ we have 
$P(\tau,t)d\tau =\Psi(\phi,\eta)d\phi$ and consequently
\begin{equation}
\label{stddimless}
\Psi(\phi,\eta) = {F^{+}(\eta \phi )F^{-}(\eta (1-\phi)) 
\over \pi (\phi (1-\phi))^{1/2}} \ .
\end{equation} 

We briefly examine some limits of the STD. For small velocity, 
or small times, we have $\eta \ll 1$, and thus
\begin{equation}
\label{stdsmallv}
\Psi(\phi,\eta) = {1 \over \pi [\phi(1-\phi)]^{1/2}} - \left ( {\eta \over \pi}
\right )^{1/2} \left ( {1\over (1-\phi)^{1/2}} - {1 \over \phi^{1/2}} \right
) + \eta \left [ {1 \over \pi (\phi (1-\phi))^{1/2}} -1 \right ] + 
O(\eta ^{3/2})\ ,
\end{equation} 
where the leading term is the arcsine law of Levy, as required.

The limiting forms of $\Psi $ for $\phi \rightarrow 0$ and 
$\phi \rightarrow 1$ are given below, along with the various forms of 
large-$\eta $ behaviour: 
\begin{equation}
\label{stdlimits}
\Psi (\phi, \eta) \sim \left \{
\begin{array}{ll}
& {F^{-}(\eta)\over \pi \phi^{1/2}} \ , \hspace{1cm} \phi \rightarrow 0 \ , \\
& \\
& {F^{+}(\eta) \over \pi (1-\phi)^{1/2}} \ , \hspace{0.5cm} 
\phi \rightarrow 1 \ , \\
& \\
& {e^{-\eta \phi} \over (\pi \eta \phi^{3})^{1/2}} \ , \hspace{0.6cm} 
\eta \rightarrow \infty , \ \eta \phi \rightarrow \infty , \ \eta (1-\phi) 
\rightarrow \infty \ , \\
& \\
& 2\left ( {\eta\over \pi \phi} \right )^{1/2} \ , \hspace{0.4cm}
\eta \rightarrow \infty , \ \phi \rightarrow 0 , \ 
\eta\phi \rightarrow 0 \ , \\
& \\
& {e^{-\eta}\over 2\pi \eta (1-\phi)^{1/2}} \ , \hspace{0.3cm}
\eta \rightarrow \infty, \ \phi \rightarrow 1 , \ 
\eta (1-\phi) \rightarrow 0 \ .
\end{array}
\right. 
\end{equation}

The distribution has a single minimum whose position $\phi _{\rm min}(\eta)$
approaches unity as $\phi _{\rm min}(\eta) \sim 1-r/\eta$ for $\eta \gg 1$,
where the universal number $r$ satisfies the transcendental equation
\begin{equation}
\label{trans}
(\pi r)^{1/2}e^{r}{\rm erfc}(-r^{1/2})=(1-2r)/2r \ ,
\end{equation}
and has the value $r=0.2040539\dots $.

It is hoped that the method of derivation presented here may be of use in 
calculating STD's for other simple stochastic processes, and that the 
explicit solution given in Eq.(\ref{stdfinal2}) will find useful application 
in a range of problems. It is currently being employed in an ecological 
context, to calculate survival probabilities for organisms in the presence 
of a step-like drifting environmental boundary\cite{na}.

\vspace{0.5cm}

The author would like to thank Satya Majumdar for an enjoyable correspondence,
and for generously sharing his expert knowledge of the literature.

\appendix

\section{}

In this appendix we calculate the function $J_{n}(u)$, which satisfies the
integro-difference equation
\begin{equation}
\label{app1}
J_{n}(u) = \int \limits _{0}^{\infty} du' \ e^{-|u-u'|} J_{n-1}(u') \ ,
\end{equation}
with $J_{0}(u)=e^{-u}$. This is accomplished by means of the generating
function
\begin{equation}
\label{app2}
{\tilde J}(u,z) = \sum \limits _{n=0}^{\infty} z^{n}J_{n}(u) \ .
\end{equation}
Summing over Eq.(\ref{app1}) we find
\begin{equation}
\label{app3}
{\tilde J}(u,z) = e^{-u}+z\int \limits _{0}^{\infty}du' \ e^{-|u-u'|}
{\tilde J}(u',z) \ .
\end{equation}
Differentiating twice with respect to $u$ yields the differential equation
\begin{equation}
\label{app4}
\partial _{u}^{2}{\tilde J}(u,z) = (1-2z){\tilde J}(u,z) \ .
\end{equation}
Demanding that ${\tilde J}(u,z)$ is finite as $u \rightarrow \infty$ gives
\begin{equation}
\label{app5}
{\tilde J}(u,z) = \left ( {1-(1-2z)^{1/2} \over z} \right )
\exp [-(1-2z)^{1/2}u]  \ ,
\end{equation}
where the $z$-dependent prefactor is found from substitution into 
Eq.(\ref{app3}). The function $J_{n}(u)$ may be recovered from the generating
function by means of the contour integral
\begin{equation}
\label{app6}
J_{n}(u) = \oint \limits _{C} {dz \over 2\pi i} 
{{\tilde J}(u,z)\over z^{n+1}} \ ,
\end{equation}
where the circular contour $C$ runs counter-clockwise about the origin and 
excludes all singularities other than the pole at the origin.

\end{document}